# Culture and International Usability Testing: The Effects of Culture in Structured Interviews

Ravikiran Vatrapu
Communication and Information Sciences
University of Hawaii at Manoa
Honolulu, HI 96822
Email: ravikira@hawaii.edu

Manuel A. Pérez-Quiñones
Department of Computer Science
Virginia Tech
Blacksburg, VA 24061
Email: perez@vt.edu

**ABSTRACT**
The global audience for software products includes members of different countries, religions, and cultures: people who speak different languages, have different life styles, and have different perceptions and expectations of any given product. A major impediment in interface development is that there is inadequate empirical evidence for the effects of culture in the usability engineering methods used for developing user interfaces. This paper presents a controlled study investigating the effects of culture on the effectiveness of structured interviews in usability testing. The experiment consisted of usability testing of a website with two independent groups of Indian participants by two interviewers; one belonging to the Indian culture and the other to the Anglo-American culture. Participants found more usability problems and made more suggestions to an interviewer who was a member of the same (Indian) culture than to the foreign (Anglo-American) interviewer. The results of the study empirically establish that culture significantly affects the efficacy of structured interviews during international user testing. The implications of this work for usability engineering are discussed.

**ACM Classification Keywords**
H.5.2 User Interfaces, Evaluation/methodology

**INTRODUCTION**
Culture influences not only interface design but also the design methods employed in building interfaces. Usability methods might be influenced by the culture of the participants that employ them. If that is the case, usability testing will not provide accurate information when a localized product is tested using these techniques, unless we take into consideration cultural influences. When the usability methods involve human-human interaction, such as is the case with structured-interviews; then the interaction of the cultures of the two participants must be considered.

One example where two cultures interact in usability evaluation occurs in the internationalization of products. International usability testing generally involves a usability expert from one country and a local facilitator in the target country [21]. When differences in cultures exist between the usability expert and the local users, usability methods may mask the usability problems instead of discovering them. In particular, the use of structured interviews where the interviewer is from a different culture than the participants is strongly influenced by cultural factors, as we show in this paper. Prior research has found that culture affects the usability evaluation process [1, 2, 3, 5, 6, 11, 32, 24, 33] and several of its well-known techniques. Culture affects the functioning of focus groups [1], the think-aloud protocol [33], questionnaires [3] and the understanding of metaphors and interface design [5, 6]. However, the effects of culture in structured interviews when the interviewer and the participants are from different cultures until now remained unexplored.

In this paper we present a study designed to evaluate the cultural effect on structured interviews while evaluating the usability of a website. Two groups of participants, all graduate students from India, evaluated a website. One group had an interviewer from India the other group an interviewer from the US. The group with the Indian interviewer found more errors, provided more feedback about the website, and identified more culturally sensitive materials than the group with the US interviewer.

The paper structure is as follows. The next section discusses some of the literature on culture, cultural metrics, and how they were used in this study, as well as the literature of culture in HCI. This is followed by the experimental design for the study, and a presentation of the findings. The paper finishes with the implications of the results of this study for the area of usability engineering.

**RELATED WORK**

**Culture**
Culture has been defined in many different ways by different researchers. A compiled list of over 200 different definitions of culture can be found in Kroeber and Kluckhohn's book [8]. This research uses Geert Hofstede's definition of culture: "*Culture is the collective programming of the mind which distinguishes the members of one group or category of people from another*" [8]. Hofstede's cultural model is well suited for empirical research as scores for each individual member of the culture can be computed unambiguously. Many other cultural models exist in the literature. Most of these are typologies, which are problematic in empirical research as individuals rarely fall into an *ideal type*.

**Hofstede's Cultural Model**
Hofstede's cultural model consists of five dimensions. Each dimension groups together phenomena in a society that were empirically found to occur in combination. Hofstede's seminal work on cultures in organizations formulated a framework of four dimensions of culture identified across nations. Michael Harris Bond [9, 10] added the fifth dimension.

The five dimensions in Hofstede's model are: power-distance, collectivism-individualism, uncertainty avoidance, feminity-masculinity, and long-short term orientation. In our work we used power distance because of its potential effect in structured interview use (see below).

Hofstede defines power distance as "the extent to which the less powerful members of institutions and organizations within a country expect and accept the power that is distributed unequally" [13, page 28]. People in large power distance cultures are much more comfortable with a larger status differential than small power distance cultures. In large power distance cultures, there is considerable dependence of subordinates on bosses. In small power distance cultures, there is limited dependence of subordinates on bosses, and a preference for mutual consultation, teamwork [13].

The score on a scale measures power distance of a culture/country. The cultures/countries are ranked from large to small power distances. In this paper, we studied Indian participants. India is a large power distance culture, and is ranked 10/11 with a score of 77. For comparison, the US is considered a small power distance country and is ranked 38 with a score of 40. The magnitude in a particular power distance scale is usually not that important, the importance is the ranking within the scale.

Participants from a large power distance culture in a structured-interview could be influenced in several ways. Participants may consider the interviewer as a person in a position of power and thus they may react accordingly. They many not respond as freely and openly to the interviewer. They may tone down their negative comments and make comments that are more positive to the foreign interviewer to "save face" and not to appear rude.

Facing an interviewer of the same culture, might mediate the effect. The cultural common ground between the same culture interviewer and the participants will help in effective communication and in identifying culture-related usability problems and/or design issues.

**Measurements of Culture in Cross-Cultural Research**
Any research dealing with culture must carefully be able to measure it in order to study its effect. In this study, we measured the power-distance for all Indian participants using the Early/Erez power differential scale [4]. The power differential scale is similar to the power distance questionnaire used by Hofstede but is more robust and reliable [4]. The goal was to show there was no difference between the two experimental groups with respect to power-distance.

A second concern in cross-cultural research is the issue of acculturation. Acculturation is a process that occurs when two or more cultures interact. Acculturation occurs as the dominant host culture absorbs to a certain extent the minority immigrant culture [25]. In cross-cultural research, the user's perception of his/her identity is important, as it is a subjective statement of cultural character. Individuals from the minority immigrant culture with high acculturation may behave like the individuals from the dominant host culture. This becomes an external variable in cross-cultural research. We control for effects from this external variable by measuring the acculturation level of the participants belonging to the minority immigrant culture [29]. Participants with high level of acculturation can be best used as members of the dominant host culture or not included in the study [29].

In this research we used the Suinn-Lew Asian Self Identity Acculturation (SL-ASIA) scale [25] to measure the acculturation levels of the Indian participants. We chose this scale as it was specifically designed for Asians. The SL-ASIA is a multiple-choice test with 21 questions and 5 choices per question. A score is obtained by adding across the answers for all 21 questions and dividing the total value by 21. The scores range from 1.00 (Low Acculturation) to 5.00 (High Acculturation).

Cultural anthropology research has been successfully applied to the fields of advertising and management but is largely still unapplied in HCI. One article [15] clearly draws the possible connections between Hofstede's dimensions and user interfaces. In the article, Marcus and Gould discuss how the Hofstede's dimensions and the cultural implications can be observed in a sampling of websites. For example, Marcus and Gould outline the possible influences of power distance on information access, hierarchies in mental models, value given to authority and official symbols and preference for explicit vs. implicit security regulations as seen in web designs. However, there is no discussion of the possible influences of culture in usability engineering methods.

**Cross-Cultural HCI Research**
Research done by Cliff Nass [16, 17, 18, 19] in social aspects of HCI has shown that even computer-literate users tend to use social rules and display social behavior in their interactions with computers. Social behavior is strongly grounded in culture as every person carries within himself or herself patterns of thinking, feeling and potential acting. Much of this is learned during early childhood. As soon as certain patterns of thinking, feeling and acting have established themselves within a person's mind they reside there. To learn new patterns of thinking, feeling and acting one has to unlearn the old patterns, which is more difficult than learning for the first time [8]. Hofstede's takes this claim even further by indicating that some reactions are likely and understandable given one's past (i.e. its cultural affiliation).

International usability testing is inherently cross-cultural. Further, it is conducted in a social setting. In the case of structured interviews, the social context is pronounced. Thus, the thinking, feeling, perception and reactions of users during international usability testing are influenced the participant's cultural and social background.

**Culture and Interface Design**
Fernandes [7] has identified various cultural issues of nationalism, language, social context, time, and currency, units of measure, cultural values, body positions, symbols and esthetics that need to be addressed during global interface design. Russo and Boor [23] present a checklist of cross-cultural items to be considered in interface design.

Khaslavsky [12] describes the impact of culture on usability and design, presents variables useful for incorporating culture into design and various design implications and mentions issues in localization of design. Development of software for international use is mostly done by the recommended process of internationalization and localization [14, 26, 30]. Symbols, heroes, rituals values and practices are the most important manifestations of culture [8]. The cultural issues identified by Fernandes, Russo and Boor, Khaslavsky and Elnahrawy consider only the symbols and rituals of different cultures ignoring the rest of the cultural manifestations.

**Culture and Usability**
Beu et al. emphasize that explication and understanding in a foreign cultural context is only possible if there is intense cooperation between representatives of the different cultures [1]. Beu et al. report that there were problems when data from different cultural sectors of China had to be compared to draw conclusions about the design. In China despite the fact that participant profiles had been drawn up invitations had to be sent out to decision makers instead of end users. This was along the lines of the Chinese notion of hierarchy. The quality of the usability tests and the focus group discussions varied greatly depending on the discussion leader and the setting. This is explained mostly by the differences in the way people from different cultures work.

Yeo [33] describes a study conducted to examine the efficacy of the global-software development lifecycle (global-SDLC), a Western software development approach employed to derive software for the global market. The think aloud technique collected objective data. The questionnaire System Usability Scale (SUS) and the interview collected subjective data. The results of the usability evaluation were found to be inconsistent. Yeo attributes inconsistencies to the large power distance and collectivist culture of Malaysia. The author says that it would appear that the inconsistencies stem from the participants' reluctance in providing critical negative comments, because of preservation of face and respect for hierarchy.

Marcus and Gould [15] applied Hofstede's cultural dimensions to web and user-interface design. The authors mention each of Hofstede's five cultural dimensions and the aspects of design that can be influenced by that particular dimension. They present screen shots of different web sites developed in different nations and point out the cultural influences on design. The findings amplify the cultural differences but are with out empirical evidence.

Honold examined the notion of culture and its relevance to Human-Computer Interaction and discusses the theories of culture in HCI [11]. Honold found cultural influences when a washing machine developed in Germany was used in India. Honold identifies eight cultural factors that have to be taken into consideration in any investigation of the context in which the product is used: objectives of the users, characteristics of the users, environment, infrastructure, division of labor, organization of work, mental modes based on previous experience and tools.

Day and Evers [2] studied the role of culture in interface acceptance and have found existence of cultural differences in interface acceptance. They ignored the Internet and studied globally marketed software packages only. Day and Evers [3] have done an instrumentation analysis of a questionnaire for multicultural data collection. Based on the results they recommend not using unmodified questions from other studies due to the multicultural context. They discourage the use of open-ended questions involving substantial reading, substantial writing and questions that are visually not separated well. Evers et al. [5, 6] report that results from a pilot study indicate that cultural aspects led to differences in user's expectations and understanding of the website of a virtual campus.

Sears et al. [24] examined the international differences and effect of high-end graphical enhancements on the perceived usability of World Wide Web. They found significant differences between the users belonging to the two different cultures of United States of America and Switzerland. Country Teng et al. [27] have found that culture had limited impact on some specific aspects of IT decision making. Tractinsky [28] found that culture effects the users' perception of aesthetics and apparent usability.

Nielsen recommends traveling to the target country and conducting usability tests as the best choice in international usability testing. Another alternative suggested by Nielsen is to employ local staff to conduct the usability testing [21];

based on the results of our work, this might be *required* if the culturally-sensitive comments are sought. But in general, usability assessment techniques have not been carefully studied in a cross-cultural context to evaluate cultural effects on their use.

**Interviews as a Usability Testing Method**
In Rubin's model [22], interviews are used in the development stage and in the evaluation stage of usability testing. Interviews are used in the beginning of the development stage to design the questionnaire. They are used in the last stage of the evaluation to clarify user responses and collect additional information. This research explores interviews in the last stage of the evaluation. Interviews are of two types: Structured and Open-ended [31]. Structured interviews have a pre-defined set of questions. A structured approach usually provides more reliable and quantifiable data than an open-ended interview and can be designed rigorously to avoid biases in the line of questioning. We followed the standard interview guidelines of the ISSUE Usability Evaluation Guidelines [31].

**METHOD**
We conducted a two-phase experiment to explore the effects of culture in structured interviews when international user testing involves participants from one culture and an interviewer from a different culture.

Phase one gathered demographic data, the power distance score and the acculturation score for each participant. All participants were of Indian origin. Phase two consisted of the usability evaluation of a website followed by a structured interview. All the interviews were audio and video recorded. Participants were divided in two groups for phase two: one group had an Indian interviewer and the second and Anglo-American interviewer. The Indian participants represent a large power distance culture. We selected 16 participants from phase one based on the SL-ASIA scores to counterbalance for acculturation. Each experimental group then consisted of 8 participants, 7 male participants and 1 female.

**AID Website**
For the experiment, we used a website intended primarily for Indian students at Virginia Tech. The Association for India's Development (AID) is a voluntary, not-for-profit organization that supports a wide variety of social service and development projects such as literacy, health care, vocational training, women's empowerment and children's welfare. AID is registered with the US Federal Government as a non-profit charitable association under the category 501(C) (3). The local AID chapter at Virginia Tech started in January 1998 and it is a registered student organization on campus. AID-VT gets funds to support projects in India mainly through fund-raising drives (like film festivals, classical dances, music concerts) and through selling gift certificates, kurtas (Indian ethnic shirts) and calendars. A significant contribution to the AID-VT fund is also made by many generous personal donors. AID-VT has a significant presence in the Indian student community at Virginia Tech.

For the experiment, we introduced usability problems in a local copy of the AID website. Using Nielsen's Ten Usability Heuristics [20], we introduced some problems in the website that matched the heuristics. The content and information of the AID web pages was not altered, modified or enhanced. No new web page were created and none of the existing ones were removed.

The following list shows the type of usability problems we introduced to the web site:

- The home page's color background is changed from white to saffron (orange) and the navigational bar at the bottom has less links than on the original site. Saffron is a religiously sensitive color in the Indian culture, and used in the nation's flag.
- We changed the navigation bars to make them inconsistent across pages. For example, the activities page in the redesigned site does not have the "Home" link.
- Links were arranged in inconsistent order. For example, link layout ordering on the 'Contact Us' page was changed. 'Activities' link comes before the 'About AID-VT' link in the redesigned page.
- We changed the color background of some pages to culturally sensitive colors. For example, the 'Join Us' page on the redesigned site has a black background, which is considered inauspicious in Indian culture.

**Procedure**
We recruited as participants Indian graduate students from different graduate programs at Virginia Tech. All were requested to voluntarily participate in the experiment. In phase one, 25 Indian participants were given three questionnaires, the demographic questionnaire, the Power Differential Scale, and Suinn-Lew Asian Self-Identity Acculturation Scale (SL-ASIA).

For phase two we contacted 16 of the participants in phase 1, using email and telephone correspondence. The interviewers had a fixed introduction script and a copy of the guidelines for the study. They read them at least twice to understand them clearly. The interviewer greeted each participant and welcomed him/her into the usability test room. All of this was done to control the familiarity and other characteristic traits from effecting the participants' first impressions. The Indian interviewer leveraged the cultural background by referring to the home state and the Indian cultural events at Blacksburg. Participants performed the usability testing tasks on the redesigned AID website on a Windows 2000 computer using Internet Explorer 6.0.

The participants were given a written description of the five usability tasks. The participants were asked to read each task description carefully and then perform the task. We designed the five tasks to give participants experience using the site and to expose them to most of the errors we introduced on the site. The tasks used were:

- Task #1: Become a member of AID and find out the time and location of the next community service hour of AID
- Task #2: Find out the information about the current executive committee members and find out the web co-coordinator for the AID-VT chapter.
- Task #3: Know about the project co-ordinate by Mr.Sivaram Tumma. Note down his contact info.
- Task #4: Learn about the home schools project (SSGS) and contact the co-ordinator for the project.
- Task #5: Learn about the grocery certificates. Get involved in the program.

After the completion of all the tasks, the interviewer conducted a structured interview. The interviewer took notes of the interview. The structured interview questions were centered on the five test tasks of the study. A meta-evaluator took notes in the adjoining observation room of the usability-lab. The transcripts from the interviews were made from the audio-video recordings and by cross checking with the notes.

The audio video recording of each interview was transcribed. The coding rules applied were to remove pauses like "um" and "ahh". "yeah" was interpreted as "yes". The scoring from the transcripts was independently verified with four graduate students who have taken at least the CS 5714 Usability Engineering course at Virginia Tech. All four coders were given the transcripts and the definitions of the terms. They were asked to count the total number of replies, the usability problems identified, suggestion given, positive and negative comments and any comments related to cultural issues in the interface. The definition of each of these categories for codification of the transcripts was:

- Usability Problems: interaction design flaw or a user difficulty directly associated with an interaction design flaw.
- Suggestion: subjective preference of the participant to the implemented design choice/tradeoff.
- Positive comment: participant's subjective approval of a design choice/ tradeoff.
- Negative comment: participant's subjective disapproval of a design choice/ tradeoff.
- Culture related comment: participant's reference to his/her native culture, country, customs, symbols, rituals and tradition.

There was some confusion among the coders about how to count negative comments, as they could count towards usability problems identified as well as negative comments. The rule applied was that any negative comment made about the interaction design and/or interface design element was also a usability problem found. Once this was made unambiguous, the scores from all the coders agreed in the numbers.

The independent variable of the experiment was the cultural profile of the interviewer. The dependent variables in the experiment were number of usability problems found, suggestions made, positive comments made, negative comments made, culturally related comments made, and the rating given to the interface.

The hypothesis for the experiment can be summarized as follows. We expected that participants with an Indian interviewer would provide more comments, identify more usability problems, make more negative comments, and make more culturally-sensitive comments than those with an Anglo-American interviewer. Similarly, we expected that participants would make more positive comments to the Anglo-American participant. In general, we expected participants to give a higher rating of the interface to the Anglo-American interviewer than to the Indian interviewer.

## DATA ANALYSIS AND RESULTS

The average age of the participants was 24 years old. Of the 25 participants, 22 (88%) were male. The average stay in India for all participants was 22.16 years. The average stay in US was 20.24 months with a low of 6 months and a high of 30 months. The participants belong to seven different states in India and speak seven different languages. Only 5 of the 25 participants have taken the graduate Usability Engineering class in the CS department at Virginia Tech. 9 of the 25 participants have participated in usability experiments before.

**Cultural Metrics of Participants**

Power distance score of 25 participants in phase 1 averaged 19.56 (medium-to-high range) in the power differential scale [4]. None of the participants had the maximum possible power distance score of 40 nor the minimum possible score of 5. Only 6 were in the low power distance range (5-16).

Acculturation was low (avg of 2.11) and none of the participants could be classified as bicultural or Anglo-American acculturated according to the rules of interpretation of scores given in the SL-ASIA scale [25]. The low acculturation scores mean that the influence of the majority host culture of US is not significant and the participants are representatives of the Indian culture.

For phase 2, we selected 16 participants and divided them into two groups for the usability evaluation part of the experiment. Both groups had an average power distance of 19.375. An ANOVA analysis with respect to power distance scores found no significant difference among the two groups.

The acculturation across the two groups was evenly distributed too. The group with the Indian interviewer had an average acculturation of 2.08 and the group with the Anglo-American interviewer was 2.19. An ANOVA analysis of acculturation scores found no significant difference among the two groups.

**Results of Usability Evaluation**

The dependent variables produced statistically significant results for all the analysis done. The results are shown in Table 1. Figure 1 and 2 show bar graphs depicting all the results for all dependent variables.

| Dependent variables | ANOVA two-factor without replication |
|---|---|
| Number of usability problems found | $F(1,7) = 36.75, p < 0.001$ |
| Number of suggestions made | $F(1,7) = 7.91, p < 0.03$ |
| Number of positive comments | $F(1,7) = 8.75, p < 0.03$ |
| Number of negative comments | $F(1,7) = 22.90, p < 0.003$ |
| Number of culture-related comments | $F(1,7) = 5.64, p < 0.05$ |
| Website rating given | $F(1,7) = 13.23, p < 0.01$ |

**Table 1. Statistical Results from the study**

The participants found more usability problems and made more suggestions with the Indian interviewer than with the Anglo-American interviewer. More positive comments and fewer negative comments were made by the participants to the Anglo-American interviewer, possibly leading to a false picture of subjective preferences. More importantly the participants were reluctant to make culture related comments to the Anglo-American interviewer. The whole purpose of finding culture related data from the structured interviews can be lost if a foreign culture interviewer is used. On the other hand, with the interviewer from the same culture, participants were more forthcoming.

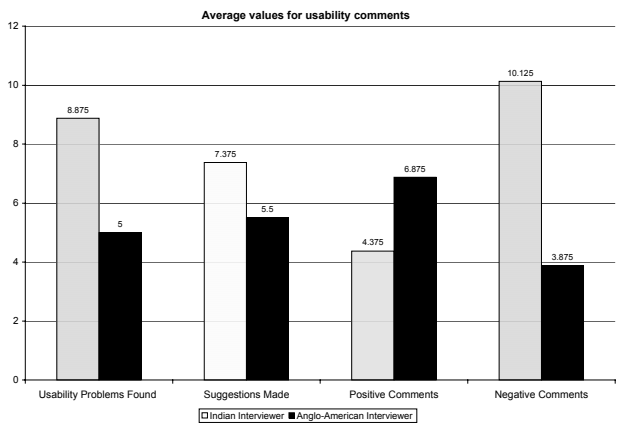

**Figure 1. Usability comments provided**

Even the rating given to the website was significantly different within each group. When asked to rate the site, with 1 being "worst" and 5 being "best" the answers came on average as shown in Figure 2. The group with the Indian interviewer gave a lower evaluation than the group with the Anglo-American interviewer.

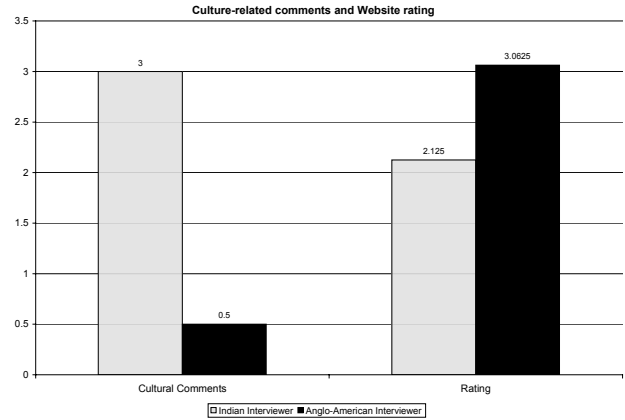

**Figure 2. Culture Related Comments and Website Rating**

**Other Observations From the Data**

We found interesting the expressiveness of some of the comments made. For example, question 2 asked the participants "What do you think about the colors used?" The responses were markedly influenced by the interviewer. Remember that we changed the background color of the home page to use saffron, one of the colors in the Indian flag.

Some comments to this question from participants with the Indian interviewer include:

#2: Indian flag colors or some thing like that are good.

#3: Ok. Since I am an Indian I know that the saffron is a color of the flag.

#4: I am not sure if they used saffron for the Indian flag or as a Virginia Tech color

The responses to the Anglo-American interviewer, were more generic and certainly lacked references to the cultural meaning of the colors. For example:

#1: … But orange color is not good.

#4: Colors are good.

#5: Orange color's purpose is not evident.

#6: Saffron and white are ok

#8: Colors are looking good.

Notice how the "feelings" about the saffron color come through in both groups, but in a culturally neutral way with the Anglo-American interviewer. In the second case, notice that the comments do not identify why the color might or might not be appropriate. Even the name of the color, "saffron" is more common in the Indian-interviewer group and "orange" in the other.

Question 18 asked about the appeal of the AID site design to the Indian students at the university. In general, we noticed that the participants used "we" to refer to the cultural group when they had an Indian interviewer. One participant referred to the collective design of the site by indicating "...we want it to be better than any site". It is interesting to contrast this statement with a comment made by one participant to the Anglo-American. The participant referred to the designers of the site in third person ("They should focus more on Indian students").

One final set of comment is worth pointing out. The original site of the AID organization has a logo that includes a drawing of Mahatma Gandhi. No participants with the Anglo-American interviewer made a reference to this image. Two participants in the other group made references to this image ("Father of Nation's image is good" and "I like Gandhi's image"). It is this type of culturally-specific comments that international usability testing is intended to obtained. Yet, by ignoring the influence of culture in the usability evaluation methods, usability engineers are bound to miss identifying these culture-specific usability issues.

## DISCUSSION

The results shown in this paper have a serious impact on the process of developing international user interfaces. In particular, the results have implications to international usability testing with users from one culture and the interviewer from a different culture. The results empirically establish that culture affects the type of responses participants provided in a structured interview. Participants responded more freely and accurately to the interviewer from the same culture than to the interviewer from a different culture.

In our particular study, the users helping in the evaluation of the interface were from a large-power distance culture. We believe that this might have had a lot to do with the results obtained. It is difficult, however, to evaluate all possible combinations of two cultures to effectively determine which factor might have had a stronger impact. A question that should be asked next is, what happens if we repeat the experiment with Anglo-American users and with two interviewers from two different cultures. Would the results be the same? We did not pursue this option because it did not represent a realistic situation in our world today. There is more software developed in the Anglo-American world and then localized to other countries than the other way around. But from a research point of view, it might be worth considering the question and its possible implications.

A more relevant question is the effect that differences in cultures might have in other usability evaluation techniques. We have shown one case where difference in culture affects the results of the evaluation. In our particular instance, the usability method, structured-interview, depends heavily on human-human interaction. This gives the opportunity for social and cultural norms and practices to come to the front. But, would the same be true in other usability evaluation methods? For example, how about if participants were doing a heuristic evaluation and they were asked to simply write down the problems found. Maybe such a method would not have as much influence from social and cultural factors. But this needs to be determined via more research.

Our initial plan was to have a third group. In this group the interviewer would be an Anglo-American, but one that would show vast knowledge of the Indian culture. From the graduate students pool at our university, we were not able to find an Anglo-American that had visited India or that had extensive knowledge of India's culture. We did not feel that "coaching" this interviewer would have been sufficient to show knowledge of the culture. Nevertheless, we wonder if showing sensitivity to the cultural issues and values might be enough to close the gap that was found in the experiment.

Finally, the research methodology followed can be employed to design and evaluate other cross-cultural usability evaluations. Successful integration of Hofstede's cultural model and acculturation with usability engineering will result in exciting and useful results in cross-cultural HCI research. Our work serves as an example of the use of cultural dimension and acculturation models in HCI research.

## ACKNOWLEGEMENTS
This research was supported by NSF Career award IIS-0049075. The authors would like to thank Raquel Ferreira for the research collaboration that helped develop this research topic.